\documentstyle[aps,preprint]{revtex}

\begin{document}
\preprint{
SNUTP 96069;~
YUMS 96-15}

\title{The Effect of Brick Walls on the
Black Hole Radiation}

\author{Sang Pyo Kim$^a$ \footnote{Electronic
mail: sangkim@knusun1.kunsan.ac.kr},
Sung Ku Kim$^b$ \footnote{Electronic mail:
skkim@theory.ewha.ac.kr},
Kwang-Sup Soh$^c$ \footnote{Electronic mail:
kssoh@phyb.snu.ac.kr},
and Jae Hyung Yee$^d$ \footnote{Electronic mail:
jhyee@phya.yonsei.ac.kr}}

\address{$^a$ Department of Physics,
Kunsan National University,
Kunsan 573-701, Korea \\
$^b$ Department of Physics,
Ewha Womans University,
Seoul 120-750, Korea\\
$^c$ Department of Physics Education,
Seoul National University,
Seoul 151-742, Korea\\
$^d$ Department of Physics
and Institute for Mathematical
Sciences,
Yonsei University, \\
Seoul 120-749, Korea}

\maketitle
\begin{abstract}
In order to understand the physical effect of
the brick wall boundary condition,
we compute the distribution of the zero-point energy
of the massless scalar fields minimally coupled to
the Schwarzschild and Reissner-Nordstr\"{o}m
black hole backgrounds.
We find that the black hole radiation
spectrum depends on the positions of the brick wall
and the observer, and reveals the interference
effect due to the reflected field by the brick wall.
\end{abstract}

Since the Bekenstein's introduction of the idea of
black hole entropy~\cite{bekenstein} and the subsequent
discovery of the black hole radiation
phenomena by Hawking \cite{hawking},
many attempts have been made to understand the
origin of the black hole entropy.
One of such proposals was 't~Hooft's brick wall
near the horizon, across which no quantum fields
can propagate~\cite{thooft}. Recently the brick
wall method has been successfully utilized in
understanding the divergence structure of
the entropy contribution of quantum fields propagating
outside the black hole horizon~\cite{alwis},
and the fact that such divergences can be absorbed
as the renormalization of the gravitational
and cosmological constants~\cite{demers,kim}.

Although the introduction of brick wall
enabled us to understand many aspects of black hole
entropy, it raises a puzzling question as to
whether it really gives the correct physics
since the existence of brick wall prohibits
the propagation of matter across the horizon
whereas it is known that black holes can absorb
matter. To understand this question, in this letter,
we compute the radiation spectra of black holes
with brick wall boundary conditions imposed.

There exist many different methods to evaluate
the radiation spectrum of black hole background~\cite{birrel}.
For the purpose of understanding the effect of brick walls
on black hole radiation,
it is convenient to use the interpretation
that the gravitational field of black hole deforms the
energy spectrum of the zero-point field and makes it appear
as a thermal radiation spectrum~\cite{hacyan,kim2}.

We consider a massless scalar field $\phi(x)$
minimally coupled to a black hole background. Then
the vacuum expectation value of energy density
observed by an observer at $x^\mu = x^\mu (\tau)$
is given by
\begin{equation}
{\cal E} = \frac{1}{2} \bigl(\xi^\mu \xi_\mu \bigr)^{1/2}
\Bigl< - \phi \frac{d^2 \phi}{d\tau^2}
- \frac{d^2 \phi}{d\tau^2} \phi + 2 \frac{d \phi}{d \tau}
\frac{d \phi}{d \tau} \Bigr>,
\end{equation}
where $\tau$ is the proper time  of the observer
and $\xi^\mu$ is the time-like Killing vector.
Using the definition of the Wightman functions
$D^{\pm} (x^\mu, x'^{\mu})$, evaluated at two points
$x^\mu = x^\mu (\tau + \sigma/2)$ and
$x'^\mu = x^\mu (\tau - \sigma/2)$ along the world-line
of the observer, the
vacuum energy density can be written as~\cite{hacyan}
\begin{equation}
{\cal E} = \frac{1}{\pi} \bigl(\xi^\mu \xi_\mu \bigr)^{1/2}
\int_0^\infty d\omega \omega^2 \bigl[\tilde{D}^+ (\omega, \tau)
+ \tilde{D}^- (\omega, \tau) \bigr],
\label{vac en}
\end{equation}
where the Fourier transforms of the Wightman functions
are defined by
\begin{equation}
\tilde{D}^{\pm} (\omega, \tau) = \int_{- \infty}^{\infty}
d\sigma e^{i \omega \sigma} D^{\pm} (\tau + \sigma/2,
\tau - \sigma/2).
\end{equation}
The distortion of the zero-point energy (i.e,
the Hawking radiation of black hole) can then be evaluated
by Eq. (\ref{vac en}) from the appropriate Wightman
functions
for the scalar field in the black hole background spacetime.

We first consider the Schwarzschild black hole
\begin{equation}
ds^2 = - \frac{2M}{r} e^{-r/2M}d\overline{u} d\overline{v}
\label{sch bl}
\end{equation}
where $M$ is the black hole mass and the Kruskal
coordinates are defined as
\begin{eqnarray}
\overline{u} &=& - 4M e^{-u/4M},~~~u = t - r^*,
\nonumber\\
\overline{v} &=&  4M e^{v/4M},~~~v = t + r^*,
\end{eqnarray}
where $r^*$ is defined by $dr^* = dr/\bigl(1 - 2M/r\bigr)$.
Here we only consider the radial motion
and the system to be 2-dimensional. The metric (\ref{sch bl})
admits a time-like Killing vector $\xi^\mu$ with magnitude
$\bigl(\xi^\mu \xi_\mu \bigr)^{1/2} = \bigl(1 - 2M/r \bigr)^{1/2}$,
and the world-line of a detector at $r=r_0$ can be represented
by
\begin{eqnarray}
\overline{u} &=& - \frac{1}{b} e^{- a \tau + b r^*_0}
\nonumber\\
\overline{v} &=& \frac{1}{b} e^{a \tau + b r^*_0}
\label{kr co}
\end{eqnarray}
where $\tau$ is the proper time of the detector,
$r^*_0$ is the position of the detector in tortoise
coordinates and
\begin{eqnarray}
a &=& \frac{\Bigl(1 - \frac{2M}{r_0} \Bigr)^{-1/2}}{4M}
\nonumber\\
b &=& \frac{1}{4M}.
\end{eqnarray}
The Wightman functions in the open $(\overline{u},
\overline{v})$ space are given by~\cite{birrel,hacyan}
\begin{equation}
D^{\pm} (\overline{u}, \overline{v}, \overline{u'},
\overline{v'}) = - \frac{1}{8 \pi}
\ln \bigl[(\overline{u} - \overline{u'} \mp
i \epsilon) (\overline{v} -\overline{v'} \mp
i \epsilon ) \bigr].
\end{equation}

't Hooft's brick wall method consists of introducing
a boundary condition
\begin{equation}
\phi (x) = 0,~~~ {\rm for}~~~ r \leq 2M + h
\end{equation}
where $h << 2M$. Thus with the brick wall boundary
condition, the Wightman functions must satisfy the conditions
\begin{equation}
D^{\pm} (x,x') \vert_{r = 2M+h} = 0 =
D^{\pm} (x,x') \vert_{r' = 2M+h}.
\label{b c}
\end{equation}
By using the image method we find the correct
Wightman functions $D^{\pm}_h (x,x')$ satisfying the brick
wall boundary condition (\ref{b c}),
\begin{eqnarray}
D^{\pm}_h (\overline{u},\overline{v})
=
&&- \frac{1}{8\pi}
\ln \Bigl[\bigl(\overline{u} - \overline{u'}
\mp i \epsilon \bigr) \bigl(\overline{v}- \overline{v'}
\mp i \epsilon \bigr) \Bigr]
\nonumber\\
&&+ \frac{1}{8\pi}
\ln
\Bigl[\bigl(- \frac{1}{b^2 \overline{v}}
e^{2 b r^*_h}  - \overline{u'}
\mp i \epsilon \bigr)
\bigl(- \frac{1}{b^2 \overline{u}}
e^{2 b r^*_h}  - \overline{v'}
\mp i \epsilon \bigr) \Bigr]
\nonumber\\
&&+ \frac{1}{8\pi}
\ln
\Bigl[\bigl( \overline{u} + \frac{1}{b^2 \overline{v'}}
e^{2 b r^*_h} \mp i \epsilon \bigr)
\bigl(\overline{v} + \frac{1}{b^2 \overline{u'}}
e^{2 b r^*_h}  \mp i \epsilon \bigr) \Bigr]
\nonumber\\
&&- \frac{1}{8\pi}
\ln
\Bigl[e^{4br^*_h} \bigl(- \frac{1}{b^2 \overline{v}}
+ \frac{1}{b^2 \overline{v'}}
\mp i \epsilon \bigr)
\bigl(- \frac{1}{b^2 \overline{u}}
+ \frac{1}{b^2 \overline{u'}}
\mp i \epsilon \bigr) \Bigr],
\label{wig}
\end{eqnarray}
where $r^*_h = r^* \vert_{r = 2M+h}$.
Using Eq. (\ref{kr co}) we find
\begin{eqnarray}
D^{\pm}_h (\tau + \sigma/2, \tau - \sigma/2)
&=& - \frac{1}{2 \pi}
\ln \Bigl[\frac{2}{b} e^{b r^*_h}
\sinh \bigl(\frac{a\sigma}{2} \mp i \epsilon
\bigr) \Bigr]
\nonumber\\
&&+  \frac{1}{4 \pi}
\ln \Bigl[\frac{4}{b^2} e^{2b r^*_h}
\sinh \bigl(\frac{a\sigma}{2} +b (r^*_0 - r^*_h)
 \mp i \epsilon
\bigr)
\nonumber\\
&&~~~~~~\times \sinh \bigl(\frac{a\sigma}{2}
- b( r^*_0 - r^*_h) \mp i \epsilon
\bigr) \Bigr].
\label{11}
\end{eqnarray}
The Fourier transforms of the Wightman functions
(\ref{11}) are
\begin{eqnarray}
\tilde{D}^+ (\omega, \tau)
&=& \frac{1}{\omega} \frac{e^{2\pi \omega/a}}{
e^{2 \pi \omega /a} -1} \Biggl[1 - \cos
\Bigl( \frac{\omega (r^*_0 - r^*_h)}{2 M a}
\Bigr) \Biggr],
\nonumber\\
\tilde{D}^- (\omega, \tau)
&=& \frac{1}{\omega}
\frac{1}{
e^{2 \pi \omega /a} -1} \Biggl[1 - \cos
\Bigl( \frac{\omega (r^*_0 - r^*_h)}{2 M a}
\Bigr) \Biggr],
\end{eqnarray}
for positive $\omega$. Using Eq. (\ref{vac en})
we thus find the energy density of the zero-point field,
\begin{equation}
{\cal E} = \frac{2}{\pi} \Bigl(1 - \frac{2M}{r_0} \Bigr)^{1/2}
\int_0^\infty d \omega \omega
\Bigl(\frac{1}{2} + \frac{1}{
e^{2 \pi \omega /a} -1} \Bigr) \Biggl[1 - \cos
\Bigl( \frac{\omega (r^*_0 - r^*_h)}{2 M a}
\Bigr) \Biggr].
\label{en1}
\end{equation}
As in the case of the Schwarzschild background without
a brick wall~\cite{hacyan}, we find the original energy
of the zero-point energy plus an additional Planckian
distribution term at the temperature
\begin{equation}
kT = \frac{a}{2\pi} = \frac{\Bigl(1 -
\frac{2M}{r_0}\Bigr)^{-1/2}}{8 \pi M}.
\end{equation}
The result, however, has an extra factor, the last factor
of Eq. (\ref{en1}), due to the existence of the brick wall
boundary condition which prevents matter fields from
crossing the wall. This factor depends on the positions
of the brick wall and the detector, and vanishes when
$r_0 = r_h$, i.e, when the detector is right at the brick wall.

We now consider the Reissner-Nordstr\"{o}m black hole
\begin{equation}
ds^2 = - \Bigl( 1 - \frac{r_+}{r} \Bigr)
\Bigl(1 - \frac{r_-}{r} \Bigr)e^{-2br^*}
d\overline{u} d\overline{v},
\label{rn bl}
\end{equation}
where $r^{\pm} = M \pm \sqrt{M^2 -Q^2}$,
$M$ and $Q$ are the mass and the charge of the black hole,
respectively, and the Kruskal coordinates are
defined by
\begin{eqnarray}
\overline{u} &=& - \frac{1}{b} e^{-bu} + \frac{1}{b},
~~u = t - r^*
\nonumber\\
\overline{v} &=&  \frac{1}{b} e^{bv} - \frac{1}{b},
~~v = t + r^*
\end{eqnarray}
with $r^*$ defined by $dr^* = dr/\bigl(1 - 2M/r + Q^2/r^2\bigr)$.
We also consider only the radial motion
and the system to be 2-dimensional. The metric (\ref{rn bl})
admits a time-like Killing vector $\xi^\mu$ with magnitude
$\bigl(\xi^\mu \xi_\mu \bigr)^{1/2} = \bigl(1 - 2M/r
+ Q^2/r^2\bigr)^{1/2}$,
and the world-line of a detector at $r=r_0$ is represented
by
\begin{eqnarray}
\overline{u} &=& - \frac{1}{b} e^{- a \tau + b r^*_0}
+ \frac{1}{b}
\nonumber\\
\overline{v} &=& \frac{1}{b} e^{a \tau + b r^*_0}
- \frac{1}{b}
\label{kr co2}
\end{eqnarray}
where $r^*_0$ is the position of the detector in
tortoise coordinates and
\begin{eqnarray}
b &=& \frac{r_+ - r_-}{2r^2_+},
\nonumber\\
a &=& b \Bigl(1 - \frac{2M}{r_0}
+ \frac{Q^2}{r^2_0} \Bigr)^{-1/2}.
\label{con1}
\end{eqnarray}
The Wightman functions $\Delta^{\pm}_h (x,x')$
satisfying the brick wall boundary conditions
\begin{equation}
{\Delta}^{\pm}_h (x,x') \vert_{r = r_+ + h} = 0 =
{\Delta}^{\pm}_h (x,x') \vert_{r' = r_+ +h}
\label{b c1}
\end{equation}
are given by
\begin{eqnarray}
{\Delta}^{\pm}_h (\overline{u},\overline{u'};\overline{v},
\overline{v'})
=
&& - \frac{1}{8\pi}
\ln \Bigl[\bigl(\overline{u} - \overline{u'}
\mp i \epsilon \bigr) \bigl(\overline{v}- \overline{v'}
\mp i \epsilon \bigr) \Bigr]
\nonumber\\
&&+ \frac{1}{8\pi}
\ln
\Bigl[- \frac{1}{b^2
\bigl(\overline{v} + \frac{1}{b} \bigr)}
e^{2 b r^*_h} - \overline{u'}
+ \frac{1}{b}
\mp i \epsilon \Bigr]
\Bigl[- \frac{1}{b^2 \bigl(\overline{u}
- \frac{1}{b} \bigr)}
e^{2 b r^*_h} - \overline{v'}
- \frac{1}{b}
\mp i \epsilon \Bigr]
\nonumber\\
&&+ \frac{1}{8\pi}
\ln
\Bigl[ \overline{u} - \frac{1}{b}
+ \frac{1}{b^2 \bigl(\overline{v}
+ \frac{1}{b} \bigr)}
e^{2 b r^*_h} \mp i \epsilon \Bigr]
\Bigl[\overline{v} + \frac{1}{b}
+ \frac{1}{b^2 \bigl(\overline{u} -
\frac{1}{b} \bigr)}
e^{2 b r^*_h}  \mp i \epsilon \Bigr]
\nonumber\\
&&- \frac{1}{8\pi}
\ln
\Bigl[e^{4br^*_h} \Bigl(- \frac{1}{b^2
\bigl(\overline{v} + \frac{1}{b} \bigr)}
+ \frac{1}{b^2 \bigl(\overline{v'}
+ \frac{1}{b} \bigr)}
\mp i \epsilon \Bigr)
\nonumber\\
&&~~~~~~\times
\Bigl(- \frac{1}{b^2 \bigl( \overline{u} -
\frac{1}{b} \bigr)}
+ \frac{1}{b^2 \bigl(\overline{u'}
- \frac{1}{b} \bigr)}
\mp i \epsilon \Bigr) \Bigr]
\label{wig1}
\end{eqnarray}
where $r^*_h = r^* \vert_{r = r_+ +h}$.
This result is exactly the same as the Schwarzschild
case except for the fact that $\overline{u}$
and $\overline{v}$ of Eq. (\ref{wig}) are
replaced by $\overline{u} - 1/b$ and $\overline{v}
+ 1/b$, respectively, and the constants
$a$ and $b$ are defined now
by Eq. (\ref{con1}). We thus obtain the energy density
in the Reissner-Nordstr\"{o}m black hole background,
\begin{equation}
{\cal E} = \frac{2}{\pi} \Bigl(1 - \frac{2M}{r_0}
+ \frac{Q^2}{r^2_0} \Bigr)^{1/2}
\int_0^\infty d \omega \omega
\Bigl(\frac{1}{2} + \frac{1}{
e^{2 \pi \omega /a} -1} \Bigr) \Biggl[1 - \cos
\Bigl( \frac{2 \omega b(r^*_0 - r^*_h)}{ a}
\Bigr) \Biggr].
\label{en2}
\end{equation}
This implies that the black hole radiates
at the temperature
\begin{equation}
kT = \frac{a}{2\pi} =
\frac{r_+ - r_-}{4 \pi r_+^2}
\Bigl(1 -
\frac{2M}{r_0} + \frac{Q^2}{r^2_0} \Bigr)^{-1/2}
\end{equation}
which is consistent with that of other methods~\cite{gosh},
but the radiation spectrum is modified due to the existence
of the  brick wall boundary condition as in the case of
the Schwarzschild black hole.

We have shown that, for both the Schwarzschild and
Reissner-Nordstr\"{o}m  black holes, the radiation
spectra are drastically modified by the brick wall
boundary conditions.
The brick wall provides a mirror-like boundary condition, and
there appear reflected fields. The cosine terms in the
last factors (\ref{en1}) and (\ref{en2}) represent the
interference effects due to the reflected field by the brick wall.
Therefore the introduction of the brick wall deforms the
local distribution of the energy spectra, even though
the overall thermodynamic entropy may look similar to
the one without a brick wall.
This implies therefore that, although the brick wall
method enabled us to clarify many aspects of black hole
entropy, the presence of brick wall near horizon
may considerably modify other aspects of black hole physics.

\section*{acknowledgments}
This work was supported in part by the Korea Science and
Engineering Foundation under Grant No. 951-0207-56-2,
95-0701-04-01-3,
and in part by the Basic Science Research Institute Program,
Ministry of Education
under Project No. BSRI-96-2418, BSRI-96-2425, BSRI-96-2427,
and by the Center for Theoretical
Physics, Seoul National University.
SKK was also supported in part by Non-Directed
Research Fund of the Korea Research Foundation.

\end{document}